\documentclass[journal=nalefd,manuscript=letter]{achemso}
\usepackage[version=3]{mhchem}
\usepackage{graphicx}
\usepackage{bm}

\author{Kristian Storm}
\affiliation{Solid State Physics/Nanometer Structure Consortium,
Lund University S-221 00 Lund, Sweden}

\author{Gustav Nylund}
\affiliation{Solid State Physics/Nanometer Structure Consortium,
Lund University S-221 00 Lund, Sweden}

\author{Lars Samuelson}
\affiliation{Solid State Physics/Nanometer Structure Consortium,
Lund University S-221 00 Lund, Sweden}

\author{Adam P. Micolich}
\email{adam.micolich@nanoelectronics.physics.unsw.edu.au}
\affiliation{Solid State Physics/Nanometer Structure Consortium,
Lund University S-221 00 Lund, Sweden} \altaffiliation{School of
Physics, University of New South Wales, Sydney NSW 2052, Australia}

\date{\today}

\title {Realizing lateral wrap-gated nanowire FETs: Controlling gate
length with chemistry rather than lithography}

\begin{document}

\begin{abstract}
An important consideration in miniaturizing transistors is
maximizing the coupling between the gate and the semiconductor
channel.  A nanowire with a coaxial metal gate provides optimal
gate-channel coupling, but has only been realized for vertically
oriented nanowire transistors.  We report a method for producing
laterally oriented wrap-gated nanowire field-effect transistors that
provides exquisite control over the gate length via a single wet
etch step, eliminating the need for additional lithography beyond
that required to define the source/drain contacts and gate lead. It
allows the contacts and nanowire segments extending beyond the
wrap-gate to be controlled independently by biasing the doped
substrate, significantly improving the sub-threshold electrical
characteristics. Our devices provide stronger, more symmetric gating
of the nanowire, operate at temperatures between 300 to 4 Kelvin,
and offer new opportunities in applications ranging from studies of
one-dimensional quantum transport through to chemical and biological
sensing.


\end{abstract}
\maketitle

Self-assembled semiconductor nanowires offer great promise as
one-dimensional elements for a range of future electronic device
applications, from next-generation computer
technologies~\cite{ThelanderMT06,FerrySci08} to energy generation
and biological/chemical sensing.~\cite{YangNL10,GaoNL10} Vertically
oriented nanowire field-effect transistors have received
considerable attention due to the possibility of producing high
density arrays,~\cite{ThelanderMT06, NgNL04, BryllertEDL06,
GoldbergerNL06, TanakaAPEX10} but the fabrication is challenging,
involving many process steps to define the contacts and
gates.~\cite{ThelanderTED08} Fundamental studies of transport in
nanowire-based quantum devices have instead focused on nanowires
oriented laterally on the substrate, mostly due to the relative ease
of fabrication. However, the common routes to gating laterally
oriented nanowires -- namely using the
substrate,~\cite{DuanNat01,BjorkNL02} insulated metal gates
deposited directly underneath the nanowire,~\cite{FasthNL05} or
gates deposited over an oxide-coated nanowire (omega
gates)~\cite{PfundAPL06, DayehAPL07} -- make calculation of the
gate-channel capacitance difficult and lead to substantial charge
inhomogeneity within the nanowire.~\cite{KhanalNL07} Since the ideal
gate configuration for a nanowire field-effect transistor (NW-FET)
is a coaxial `wrap-gate', the development of lateral wrap-gate
NW-FETs is highly desirable. It is a challenging goal, because
depositing gate metal underneath a nanowire already sitting
laterally on a substrate is a formidable undertaking, and using a
nanowire where the wrap-gate exists prior to deposition onto a
substrate entails the difficulty of exposing the ends of the
nanowire to make contacts that are not electrically shorted to the
wrap-gate.

We have developed a surprisingly simple route to producing lateral
wrap-gated NW-FETs that builds upon existing methods for fabricating
vertical wrap-gate NW-FETs~\cite{ThelanderMT06} and exploits the
fortuitous tendency for etchants to undercut a resist, gaining
access to the wrap-gate via the openings made for the source and
drain contacts using electron-beam lithography. A major advantage is
that we can precisely control the gate length over a wide range via
a single wet-etch step. Our method requires no additional
lithography compared to the current state-of-the-art for gating
lateral NW-FETs,~\cite{PfundAPL06, DayehAPL07} and results in
NW-FETs with improved sub-threshold characteristics.

\begin{figure}
\includegraphics[width=16cm]{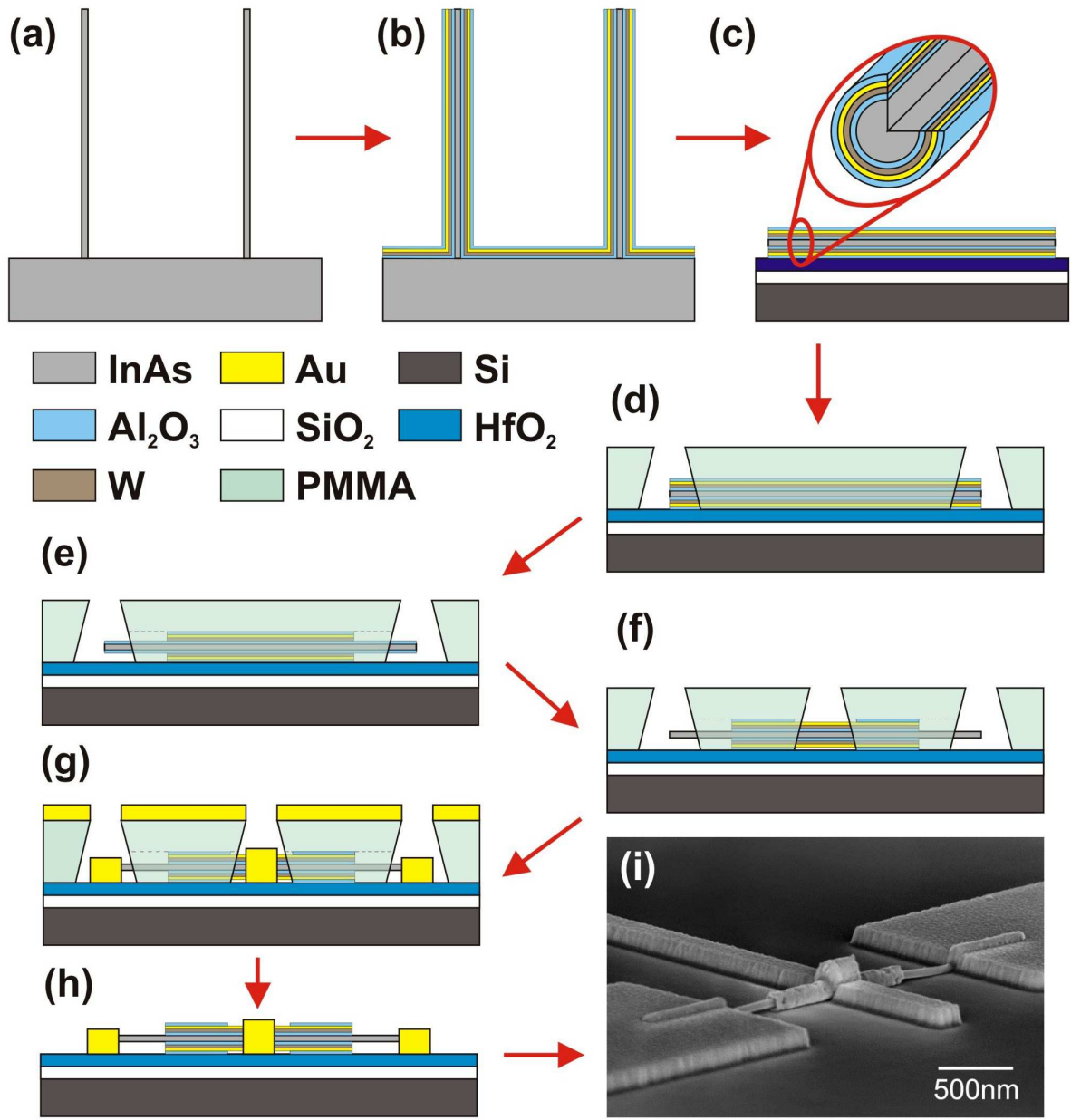}
\caption{Fabrication process for lateral wrap-gate NW-FETs, which
involves: (a) nanowire growth; (b) deposition of gate insulators and
gate metal; (c) deposition of coated nanowires onto a device
substrate (inset shows a magnified 3D section of the coated
nanowire); (d) deposition of PMMA resist and definition of source
and drain contacts by electron beam lithography (EBL); (e)
shortening of the outer oxide and wrap-gate by wet etching; (f)
definition of the gate lead; (g) Ni/Au metallization by thermal
evaporation; and (h) lift-off to give the completed structure shown
in the SEM image in (i).  The scale bar in (i) represents a length
of $500$~nm.}
\end{figure}

The process used is presented in Fig.~1. InAs nanowires $\sim 50$~nm
in diameter and $3.3~\mu$m long were grown by chemical beam epitaxy
(Fig.~1(a)).~\cite{OhlssonPhysE02} The nanowires contain a $75$~nm
long InAs$_{0.95}$P$_{0.05}$ segment, added for a separate
experiment, between two $1.6~\mu$m long InAs segments. The as-grown
nanowires were coated with $12$~nm of \ce{Al2O3} by atomic layer
deposition (ALD), a metal wrap-gate consisting of $16.5$~nm of
tungsten and $11.5$~nm of gold by DC sputtering, and a further
$12$~nm of \ce{Al2O3} as an outer insulator layer (Fig.~1(b)). This
outer oxide is superfluous electrically, but essential to the
fabrication -- we transfer the nanowires to a heavily doped Si
substrate (Fig.~1(c)) using the tip of a small piece of lab-wipe,
and the coated nanowires will not adhere to the paper fibres without
this outer oxide. The device substrate features a $30$~nm
thermally-grown \ce{SiO2} layer covered with $10$~nm of \ce{HfO2} by
ALD, and is pre-patterned with large-scale leads and alignment
markers for electron-beam lithography (EBL) prior to deposition of
the nanowires. The $10$~nm \ce{HfO2} layer is also essential to the
fabrication -- \ce{HfO2} etches much more slowly than \ce{Al2O3} and
\ce{SiO2} in buffered hydrofluoric acid (BHF), and thus the
\ce{HfO2} layer protects the \ce{SiO2} substrate insulation during
two BHF etches used to remove sections of the \ce{Al2O3} layers
coating the nanowire/wrap-gate. After nanowire deposition, the
device is coated with $\sim 400$~nm of polymethylmethacrylate (PMMA)
resist, and the source and drain leads to a selected nanowire are
defined by EBL (Fig.~1(d)).

The W/Au wrap-gate and \ce{Al2O3} layers initially extend the full
length of the nanowire, jutting out into the EBL-defined source and
drain contact regions (Fig.~1(d)). Three subsequent wet etches are
administered at this stage (Fig.~1(e)), and it is the second, an
iodine-based gold etch,~\cite{MuraMR03} that ultimately sets the
wrap-gate length. The first is a $30$~s etch in BHF at room
temperature to strip back the outer \ce{Al2O3} layer, exposing the
wrap-gate. This is followed by the Au etch, which we describe in
more detail below. The third etch in $31~\%$ \ce{H2O2} at
$40~^{\circ}$C for $70$~s removes any exposed tungsten in the
wrap-gate. The various Au etch solutions are made from a stock
solution containing $4$~g \ce{KI} and $1$~g \ce{I2} dissolved in
$80$~mL of deionized \ce{H2O} (Millipore $18$~M$\Omega$.cm). This
stock solution has an extremely high etch rate $\sim
1~\mu$m/min,~\cite{Microchem} and we use further dilutions ranging
from $1:5$ to $1:100$ in deionized \ce{H2O} to ensure that the etch
can be applied with accurate control within a reasonable etch time
($\sim 60$~s).

The lead connecting the wrap-gate is defined in a second stage of
EBL on the same resist (Fig.~1(f)), but is omitted for the devices
made to study the dependence of gate length on the etch parameters
(i.e., Figs.~2/3). A final BHF etch ($30$~s at room temperature) is
then performed to simultaneously remove the exposed inner \ce{Al2O3}
layer at the source and drain contacts and the exposed outer
\ce{Al2O3} layer where the gate lead intersects the wrap-gate
(Fig.~1(f)). Electrodes consisting of $25$~nm Ni and $75$~nm Au were
deposited by thermal evaporation in vacuum (Fig.~1(g)), immediately
following a contact passivation step involving a $120$~s immersion
in an aqueous \ce{(NH4)2S} solution at
$40~^{\circ}$C.~\cite{SuyatinNano07} Finally, lift-off of the excess
metal and PMMA is performed to obtain the completed device structure
shown schematically in Fig.~1(h). A scanning electron microscope
(SEM) image of a completed device made using the $1:20$ Au etch for
$\sim 60$~s is shown in Fig.~1(i). A corresponding top-view of a
similar device appears in the supplementary information (Fig.~S1),
and careful inspection of contrast changes along the wrap-gate
reveals where the outer oxide has been stripped to expose the
wrap-gate for contacting. The ability to define all three leads to
the nanowire, and set the gate length, using a single resist layer
and one metal evaporation makes this a very time efficient
fabrication process.

Two parameters govern the Au etch -- concentration and time -- and
in Figs.~2 and 3 we demonstrate that both provide excellent fine
control of the gate length. As the measurable parameter for gate
length we use the separation between the gate edge and the nearest
contact in the completed device, obtained from scanning electron
microscopy studies. The gate-contact separation is independent of
the nanowire length, as well as of the EBL-defined source-drain
contact separation providing that the Au etch undercut is not more
than half the source-drain contact separation. For each data point
in Fig.~2(a) and 3(a) we made a `chip' containing $24$ lateral
wrap-gate NW-FETs, each with two gate-contact separations at the
source and drain. We obtained an average yield of $72~\%$, such that
each point in Fig.~2(a) and Fig. 3(a) is the average of $30-40$
individual measurements for a given Au etch concentration/time, with
the other three wet etches held constant at the specifications given
earlier.

\begin{figure}
\includegraphics[width=8cm]{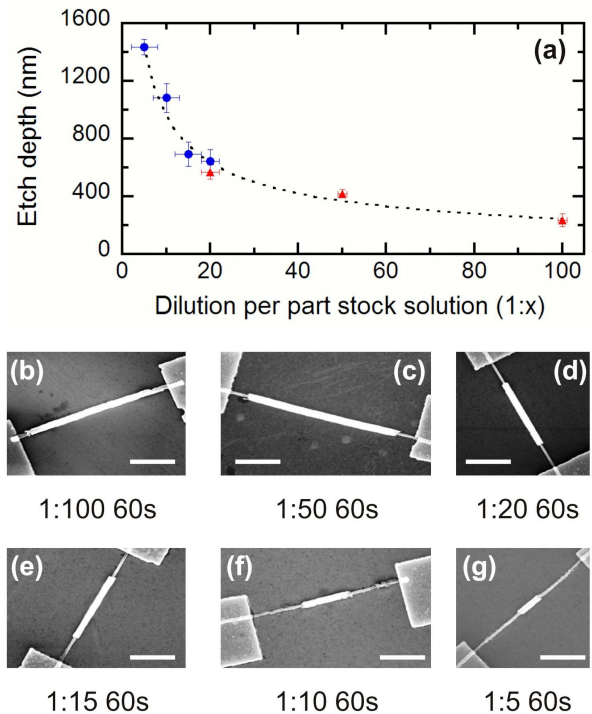}
\caption{Control of gate-length by KI/\ce{I2} etch concentration.
(a) The average gate-contact separation versus KI/\ce{I2} etchant
concentration for a constant $60$~s etch time.  The blue/red data
points were obtained from two separate process batches.  In each
case the average is taken over the yield from a chip nominally
containing $24$ devices, this yield is typically $72 \pm 17~\%$. The
vertical error bars are the corresponding standard deviation. The
horizontal error bars are an estimate based on the given process.
(b) - (g) corresponding scanning electron micrographs of a typical
device for each concentration studied. The white scale-bars indicate
a length of $1~\mu$m.}
\end{figure}

Figure~2(a) shows the resulting etch depth versus Au etch
concentration, with SEM images of representative devices for each
concentration presented in Figs.~2(b-g). In each case the Au etch
time is $60$~s. At the lowest concentration of $1:100$ stock
solution to \ce{H2O}, the wrap-gate almost spans the entire distance
between source and drain contacts (see Fig.~2(b)). A fortuitous
aspect of our method is that the undercutting action of the etch
always prevents shorting of the gate to the contacts, in contrast to
other self-aligned methods for creating quantum wires where
gate-contact shorting is a significant problem.~\cite{KaneAPL98,
SeeAPL10} Higher Au etchant concentration causes increased
under-etching at the contact openings, resulting in a wrap-gate that
is much shorter than the lithographically-defined contact
separation. At sufficiently high concentration the wrap-gate becomes
extremely short (see Fig.~2(g)), and can be eliminated entirely (not
shown), leading to a malfunctioning device if a gate lead is
deposited. The data in Fig.~2(a) roughly follows a power law
dependence, indicating an effective route to wide-ranging control
over the gate length. Varying the etch time provides an alternative
control, as demonstrated in Fig.~3(a) for the $1:100$ dilution etch,
with corresponding typical devices shown in Figs.~3(b)-(e). Although
changing the dilution and time have a similar outcome, in a
practical device fabrication setting, changes in concentration would
be more suitable for enacting coarse control over wrap-gate length,
with the etch time providing a route to finer control that enables
devices with precise gate lengths to be produced.

\begin{figure}
\includegraphics[width=8cm]{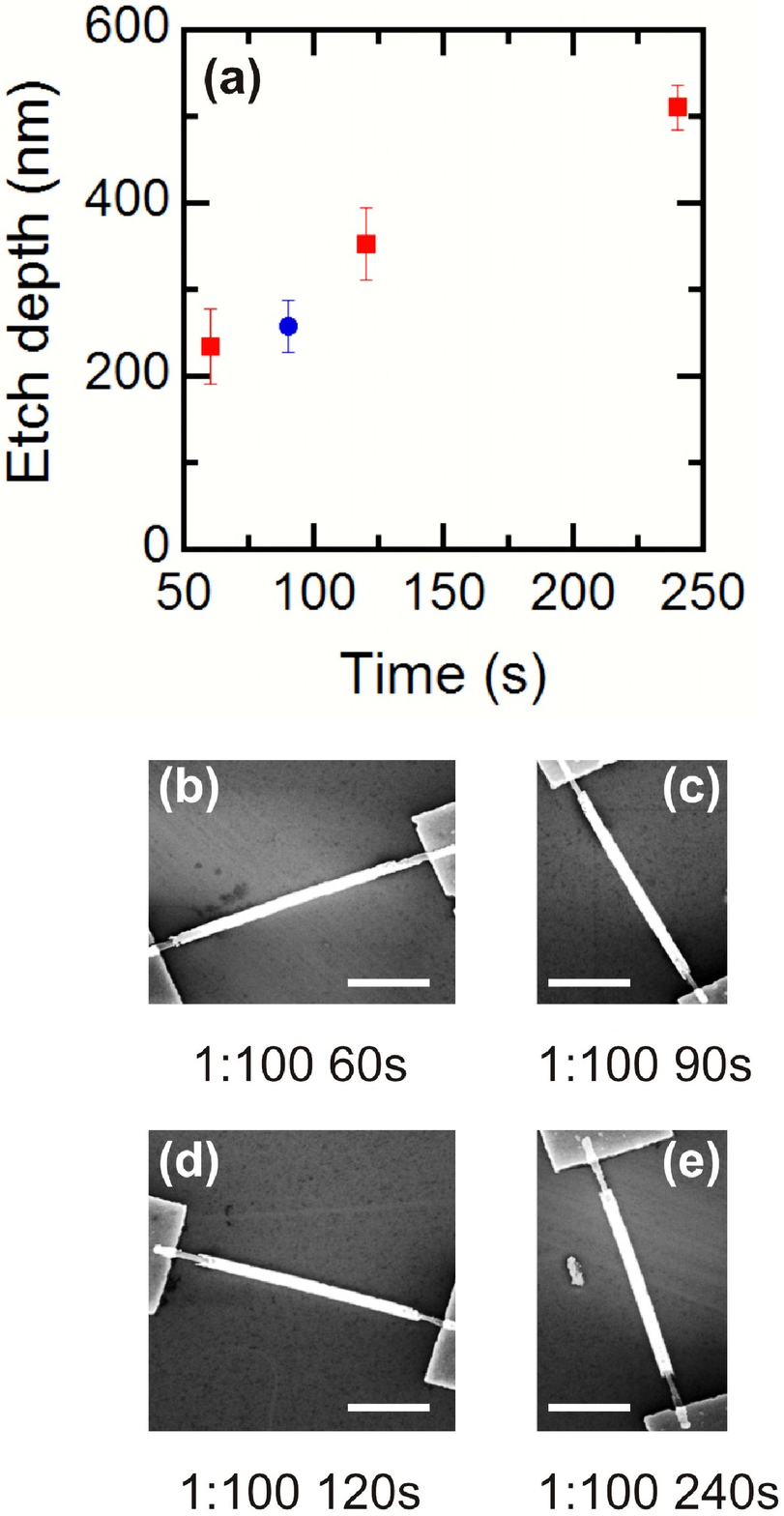}
\caption{Control of gate-length by etch time. (a) The average
gate-contact separation versus time for the $1:100$ KI/\ce{I2}
etchant. The blue/red data points were obtained from two separate
process batches. In each case the average is taken over the yield of
samples from a chip with $24$ devices on it, this yield is typically
$72 \pm 17~\%$. The vertical error bars are the corresponding
standard deviation. The horizontal error bars are an estimate based
on the given process. (b) - (e) corresponding scanning electron
micrographs of a typical device for each etch time studied. The
white scale-bars indicate a length of $1~\mu$m.}
\end{figure}

The influence that the other three wet etch processes have on the
gate length is an important additional question in light of our
findings about the Au etch. To confirm that the Au etch is indeed
the sole determinant of gate length, we made three additional chips,
each with a $1:20$ Au etch applied for $60$~s, but with one of the
three other wet-etches in the process applied for twice the time
specified earlier (i.e., BHF etches of inner/outer gate insulator
for $60$~s or \ce{H2O2} etch for W wrap-gate metal for $140$~s).
Doubling the time for any of these etches has a negligible effect on
the gate-contact separation for resulting NW-FETs (see Supplementary
Fig.~S2 for data). In contrast, doubling the time for the Au etch
increases the gate-contact separation by $40-50~\%$ based on the
data in Fig.~3. This does not mean that the specifics of these three
wet etches can be neglected, if these etches are insufficient or
overly extended then working devices are not obtained, but the
process is robust to small fluctuations in these three etches
compared to the Au etch.

\begin{figure}
\includegraphics[width=8cm]{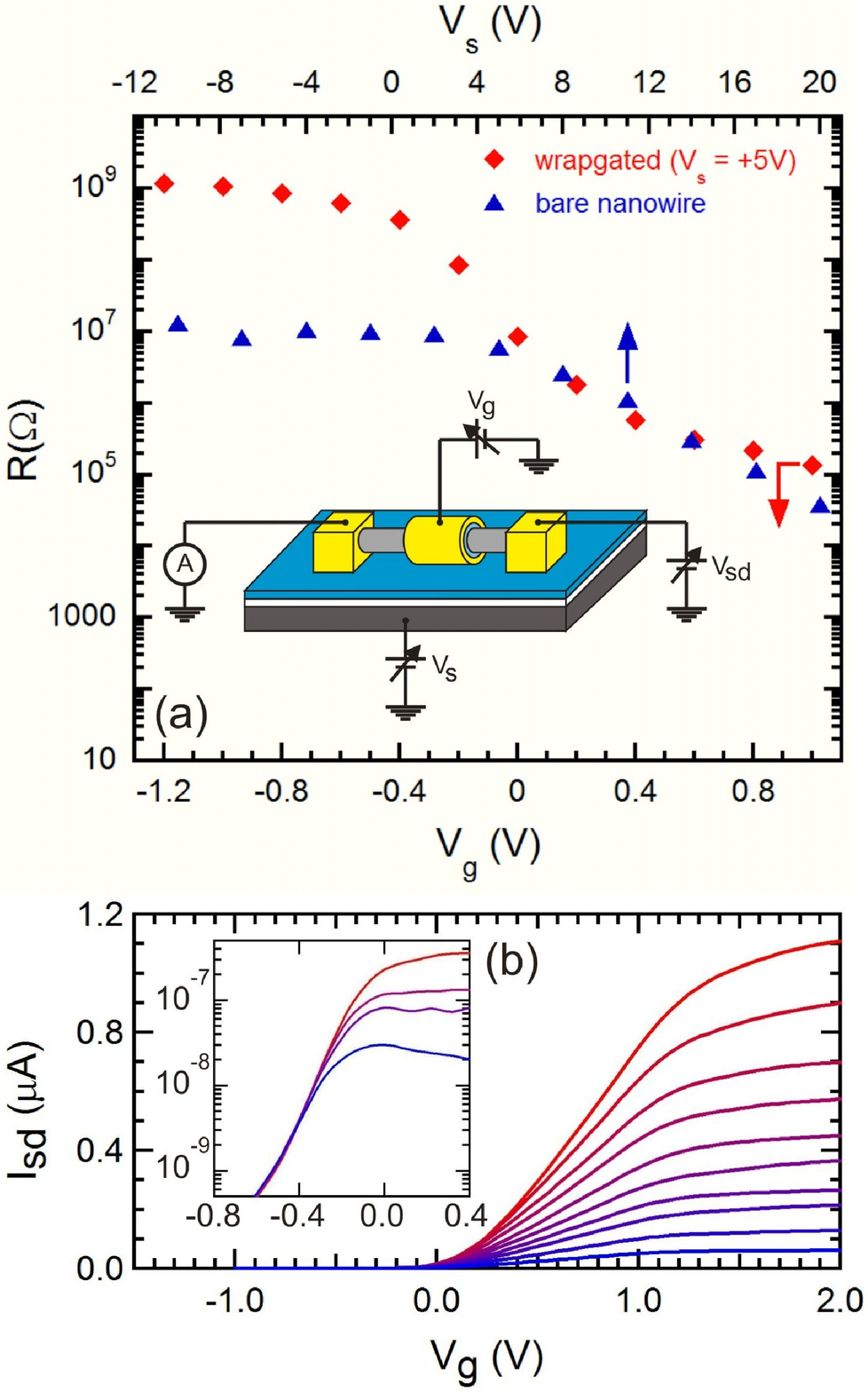}
\caption{Electrical characteristics of a lateral wrap-gate NW-FET.
(a) Device resistance $R$ versus wrap-gate voltage $V_{g}$ (red
diamonds -- bottom axis) for a wrap-gated NW-FET and substrate
voltage $V_{s}$ (blue triangles -- top axis) for a NW-FET made using
the same nanowires without wrap-gates.  The wrap-gate NW-FET data is
obtained with $V_{s} = +5$~V.  (inset) A schematic showing the
circuit used for these measurements.  For the devices without
wrap-gates, only $V_{s}$ is applied, for devices with wrap-gates,
both $V_{s}$ and $V_{g}$ are applied simultaneously.  (b)
Source-drain current $I_{sd}$ versus $V_{g}$ for source-drain biases
$V_{sd}$ from $+100$~mV (red) to $+10$~mV (blue) in steps of
$10$~mV. Data obtained at $V_{s} = +5$~V and room temperature.
(inset) $I_{sd}$ versus $V_{g}$ for different $V_{s} = 11$ (red),
$10.5$, $9.5$, $8.5$~V (blue). Data was obtained with $V_{sd} =
+50$~mV and at room temperature.}
\end{figure}

To demonstrate the electrical properties of our wrap-gate NW-FETs,
two chips A and B were prepared in separate processing runs with the
$1:20$ dilution Au etchant applied for $60$ and $100$~s,
respectively. A third chip C containing conventional substrate-gated
NW-FETs (i.e., no wrap-gate) using nanowires from the same growth
was also prepared. We used the measurement circuit shown inset to
Fig.~4(a) with the source-drain current $I_{sd}$ measured at the
drain contact as a function of biases $V_{sd}$, $V_{g}$ and $V_{s}$
applied to the source, wrap-gate and substrate, respectively. In
Fig.~4(a) we plot the measured two-terminal resistance $R$ versus
$V_{s}$ (blue triangles -- top axis) for a substrate-gated NW-FET.
The resistance increases from $38$~k$\Omega$ at $V_{s} = +20$~V,
saturating at $\sim 10$~M$\Omega$ for negative $V_{s}$. The measured
resistance for the wrap-gate NW-FET versus $V_{g}$ (red diamonds --
bottom axis) with $V_{s}$ held at $+5$~V is also shown. At $V_{g} =
+1.0$~V, $R = 137$~k$\Omega$, which compares favorably with $R \sim
5$~M$\Omega$ obtained for the substrate-gated NW-FET under
corresponding conditions (i.e., $V_{g} = +5$~V). The wrap-gate
NW-FET reaches an off-state resistance $\sim 1$~G$\Omega$, two
orders of magnitude higher than the substrate-gated NW-FET, with a
smaller gate-range, demonstrating that the wrap-gate allows much
stronger gating while freeing the substrate to be used for
maintaining the quality of the contacts.

A common quality metric for NW-FETs is the field-effect mobility
$\mu_{FE} = g_{m}L^{2}/C_{g}V_{sd}$, where $g_{m} = dI_{sd}/dV_{g}$
is the transconductance and $C_{g}$ is the gate
capacitance.~\cite{DayehSST10} Before comparing the mobility for our
devices, we note that fundamental differences in device geometry
combined with difficulties in accurately estimating the capacitance
due to the dual-layer \ce{SiO2}/\ce{HfO2} substrate insulation,
substrate gating of regions outside the wrap-gate, and the
overestimation~\cite{KhanalNL07} of the capacitance by the standard
expression for a cylindrical conductor on a planar
substrate~\cite{MorseBook53, MartelAPL98} mean that the values
obtained are semi-quantitative estimates at best. For the
substrate-gated NW-FET we obtain $\mu_{FE} = 702$~cm$^{2}$/Vs for a
contact separation of $2.4~\mu$m and gate capacitance of $415$~aF
obtained using the standard cylinder-on-plane
model.~\cite{MorseBook53, MartelAPL98} This mobility is low compared
to that typically found for substrate-gated InAs NW-FETs ($\sim 500
< \mu_{FE} < \sim 5000$~cm$^{2}$/Vs). We initially attributed this
to the InAs$_{0.95}$P$_{0.05}$ segment, however, Lind {\it et al.}
reported $\mu_{FE} \sim 1500$~cm$^{2}$/Vs for substrate-gated InAs
NW-FETs containing a $150$~nm InAs$_{0.7}$P$_{0.3}$
segment,~\cite{LindNL06} which presents a much higher transport
barrier. Hence we instead suspect that the reduced mobility is due
to surface effects, since our substrates have a \ce{HfO2} surface
layer unlike the devices studied by Lind {\it et
al.}~\cite{LindNL06} For the wrap-gate NW-FET we obtain $\mu_{FE} =
109$~cm$^{2}$/Vs for a $1.33~\mu$m channel length and $1.77$~fF gate
capacitance calculated using a coaxial capacitor model, ignoring
end-effects or capacitance contributions from the nanowire regions
extending outside the wrap-gate. The mobility is lower than that for
the substrate-gated NW-FET, and we attribute this to surface states
at the nanowire/inner oxide interface. We note that $\mu_{FE}$
values of similar magnitude were reported for vertical wrap-gated
InAs NW-FETs and likewise attributed to interface
effects.~\cite{TanakaAPEX10} Interface problems in wrap-gate NW-FETs
are a well-known issue, and their passivation is a goal for future
development of these devices.~\cite{ThelanderMT06, DayehSST10}

\begin{figure}
\includegraphics[width=8cm]{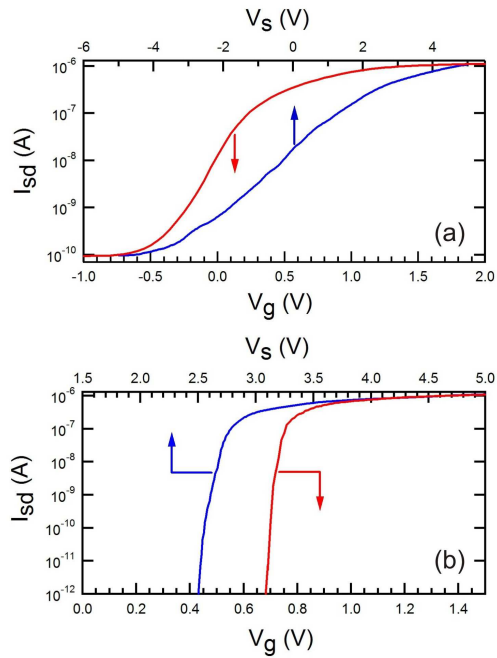}
\caption{Comparison between substrate and wrap-gate transfer
characteristics. (a) Source-drain current $I_{sd}$ vs wrap-gate
voltage $V_{g}$ at fixed substrate bias $V_{s} = +5$~V (red trace --
bottom axis) and vs $V_{s}$ at fixed $V_{g} = +2.0$~V (blue trace --
top axis) at room temperature, and (b) $I_{sd}$ vs $V_{g}$ at fixed
$V_{s} = +5$~V (red trace -- bottom axis) and vs $V_{s}$ at fixed
$V_{g} = +1.5$~V (blue trace -- top axis) at a temperature of $4$~K.
The two devices differ only in the duration of the Au etch, which
was performed for (a) $60$~s and (b) $100$~s in the $1:20$ dilution
etchant. Data was obtained with $V_{sd} = 40$~mV.}
\end{figure}

Our wrap-gate NW-FETs make up for their unexceptional mobility with
excellent sub-threshold characteristics, which are of interest for
sensing applications.~\cite{GaoNL10} Figure~4(b) shows transfer
characteristics for two wrap-gate NW-FETs with very similar
characteristics aside from a difference in threshold voltage. The
main panel of Fig.~4(b) shows plots of $I_{sd}$ versus $V_{g}$
obtained for $V_{sd}$ between $+10$ and $+100$~mV, demonstrating the
high quality characteristics afforded by using the biased substrate
to maintain stable, low resistance contacts while depleting the
nanowire using the wrap-gate.  The threshold voltage $V_{th}$ for
this device is $12$~mV but we typically find values ranging from
$-1.5$ to $0.5$~V, and a survey of the $17$ working devices on Chip
A gave an average $V_{th} = -0.67 \pm 0.34$~V. This spread in
$V_{th}$ may be due to diameter variation in the nanowires
used,~\cite{RehnstedtEL08} but may be related to charge trapping at
the InAs/\ce{Al2O3} interface, and we observe occasional discrete
shifts in $V_{th}$ consistent with this, particularly at low
temperature.

We demonstrate that the wrap-gate operates independently of the
substrate in Fig.~4(b) inset by plotting $I_{sd}$ versus $V_{g}$ for
differing $V_{s}$. When the wrap-gated region is in the on-state
(upper right), $V_{s}$ is the dominant influence over $I_{sd}$ and
small changes in $V_{g}$ have little impact. For $V_{g} < -0.2$~V,
the wrap-gate dominates, with traces for each $V_{s}$ overlapping.
To directly compare the relative effect of the wrap-gate and
substrate, in Fig.~5(a) we show the transfer characteristics for
both the wrap-gate (red trace -- bottom axis) and the substrate
(blue trace -- top axis), starting from a common configuration with
$V_{g} = +2.0$~V and $V_{s} = +5$~V. In both cases, $I_{sd}$ falls
with decreasing gate bias, saturating at an off-state value of
$100$~pA, corresponding to an on-off ratio $\sim 10^{4}$. The
substrate data is slightly noisier than the wrap-gate data for low
$I_{sd}$, as expected since substrate biasing also has a detrimental
effect on the contacts. The wrap-gate inverse sub-threshold slope is
$196$~mV/decade, nearly an order of magnitude better than the
$1480$~mV/decade obtained using the substrate. It is interesting to
compare this with results by Lind {\it et al.}~\cite{LindNL06} Their
substrate inverse sub-threshold slope decreased from $3-4$~V/decade
to $900$~mV/decade on adding the InAsP segment, and they predicted
further substantial improvement if the InAsP segment is enclosed
within a wrap-gate.~\cite{LindNL06} Our data supports this
prediction, and is consistent with recent data from a vertical
wrap-gate NW-FET with a $50$~nm InAs$_{0.8}$P$_{0.2}$ segment that
gave inverse sub-threshold slopes as low as
$120$~mV/decade.~\cite{ThelanderEDL08} On a device-to-device basis
the wrap-gate inverse sub-threshold slope is more consistent than
$V_{th}$ with an average slope of $198 \pm 38$~mV/decade across the
$17$ devices on Chip A. Although the current record for vertical
wrap-gate NW-FETs is $\sim 75$~mV/decade,~\cite{ThelanderTED08}
values from $100$ to $750$~mV/decade are
typical,~\cite{TanakaAPEX10, LindNL06, RehnstedtEL08, FrobergEDL08,
RehnstedtTED08} and heavily dependent on careful optimization of the
nanowire/insulator interface.~\cite{TanakaAPEX10,ThelanderTED08,
ThelanderEDL08, RehnstedtTED08}

The ability to achieve strong, symmetric gating of a section of
nanowire whilst using the substrate to ensure a low contact
resistance makes lateral wrap-gate NW-FETs very interesting for
studying one-dimensional quantum transport phenomena at low
temperature. To demonstrate that our devices work well at low
temperatures, in Fig.~5(b) we present transfer characteristics for
the wrap-gate with $V_{s} = +5$~V (red trace -- bottom axis) and the
substrate gate with $V_{g} = +1.5$~V (blue trace -- top axis)
obtained from Chip B at $4$~Kelvin. In both cases the inverse
sub-threshold slope has improved markedly, to $32$~mV/decade for the
wrap-gate, and $129$~mV/decade for the substrate. The off-current
has also dropped below the noise-floor $\sim 1$~pA, representing an
on-off ratio exceeding $10^{6}$.

We have reported a process for the fabrication of lateral wrap-gate
nanowire field-effect transistors. This coaxial gate architecture
holds considerable promise for studies where strong symmetric gating
is paramount. The wrap-gate length can be easily and effectively
tuned via the concentration and time for a single wet etch,
mitigating the need for additional lithography. The contacts and
nanowire segments outside the wrap-gate can be independently gated
using a doped substrate to combine stable, low resistance contacts
with the improved sub-threshold characteristics afforded by the
enhanced coupling of the wrap-gate. As with vertical wrap-gate
NW-FETs, further work is required to optimize the nanowire/insulator
interface, and would give further improvements in performance. Our
device operates at low temperature and is interesting for studies of
one-dimensional quantum transport phenomena such as spin-charge
separation~\cite{AuslaenderSci05} or Wigner
crystallization.~\cite{KristinsdottirarXiv10} Our design also
presents other opportunities, for example, omitting the gate lead
would result in a floating wrap-gate NW-FET that would combine the
high-quality sub-threshold characteristics desirable for optimal
sensitivity~\cite{GaoNL10} with an exposed gold gate surface
suitable for direct attachment of antibodies via thiol-bonding of
cysteine residues,~\cite{KaryakinAnalChem00} as well as other
gold-binding polypeptides.~\cite{BrownNatBio97}

\acknowledgement

This work was funded by the Swedish Foundation for Strategic
Research (SSF), Swedish Research Council (VR), Knut and Alice
Wallenberg Foundation (KAW) and the Australian Research Council
(ARC).  APM acknowledges an ARC Future Fellowship (FT0990285) and
thanks M.O. Williams for assistance with the gate-length dependence
study, and the Nanometer Structure Consortium at Lund University for
hospitality.  This work was performed in part using the NSW node of
the Australian National Fabrication Facility (ANFF). The authors
declare that they have no competing financial interests.

{\bf Supporting Information Available:} Extended details of methods
used, an additional SEM image, and results of a study demonstrating
the influence of the BHF and \ce{H2O2} etches on gate length. This
material is available free of charge via the Internet at
http://pubs.acs.org.

\end{document}